# Formulating Module Assessment for Improved Academic Performance Predictability in Higher Education

Mohammed Alsuwaiket
Department of Computer Science,
Loughborough University,
Loughborough, Leicestershire, UK
m.alsuwaiket@lboro.ac.uk

Anas H. Blasi
Department of Computer Information
Systems, Mutah University,
Karak, Jordan
ablasi1@mutah.edu.jo

Ra'Fat Al-Msie'deen
Department of Computer Information
Systems, Mutah University,
Karak, Jordan
rafatalmsiedeen@mutah.edu.jo

*Abstract*—**The choice of an effective student assessment method is an issue of interest in Higher Education. Various studies [1] have shown that students tend to get higher marks when assessed through coursework-based assessment methods which include either modules that are fully assessed through coursework or a mixture of coursework and examinations than assessed by examination alone. There are a large number of educational data mining (EDM) studies that pre-process data through conventional data mining processes including data preparation process, but they are using transcript data as they stand without looking at examination and coursework results weighting which could affect prediction accuracy. This paper proposes a different data preparation process through investigating more than 230,000 student records in order to prepare students' marks based on the assessment methods of enrolled modules. The data have been processed through different stages in order to extract a categorical factor through which students' module marks are refined during the data preparation process. The results of this work show that students' final marks should not be isolated from the nature of the enrolled module's assessment methods. They must rather be investigated thoroughly and considered during EDM's data pre-processing phases. More generally, it is concluded that educational data should not be prepared in the same way as other data types due to differences as data sources, applications, and types of errors in them. Therefore, an attribute, coursework assessment ratio (CAR), is proposed to be used in order to take the different modules' assessment methods into account while preparing student transcript data. The effect of CAR on prediction process using the random forest classification technique has been investigated. It is shown that considering CAR as an attribute increases the accuracy of predicting students' second-year averages based on their first-year results.**

## I. INTRODUCTION

Although educational data have been recorded and analyzed from educational software for a long time, only recently has this process been formed into a new field, educational data mining (EDM). The EDM process converts raw data from educational systems into useful information that could potentially have a great impact on educational research and practice [1]. Additionally, EDM uses a wide range of methods to analyze data, including, but not limited to, supervised and unsupervised model induction, parameter estimation, relationship mining, etc. [2, 3]. During the last few decades, the use of coursework-based module assessment increased in the UK and other countries due to various educational arguments. Additionally, it appears to be that students prefer the assessment to be based on either coursework alone or by a mix of both coursework and exams because these types of assessments tend to yield higher marks than exam-based assessment alone [4]. The increased adoption of coursework-based assessment has contributed to an increase over time in the marks on individual modules and in the proportion of good degrees across entire programmes [5]. Accurate and fair student assessment is an issue of concern in higher education (HE). Changes in the use of different assessment methods have given rise to an increasing number of universities shifting from traditional exam-based to continuous assessment throughout the semester (coursework-based) [6]. Coursework-based assessment methods differ from exam-based assessment methods where knowledge or skill is tested for a very specific period of time. Moreover, it has been widely acknowledged that the chosen assessment method will determine the style and content of student learning and skill acquisition [6]. Coursework marks are a better predictor of long-term learning of course content than are exams [7]. Nonetheless, it appears to be that none of the studies in the EDM research field reflected the assessment method used in modules on the final marks. So, this paper aims to pre-process students' transcript data differently, extract a factor from the assessment method and use it to refine student marks again to ensure more accuracy prior to processing them by using the EDM techniques.

## II. ASSESSMENT METHODS IN HIGHER EDUCATION

Regardless the inabilities to absolutely ensure student learning through different assessment methods, assessment is still an essential tool through which teachers influence the ways students respond to courses. On the other hand, there are clear steers from UK government towards coursework-based

Corresponding author: Anas H. Blasi

assessment focused on employability that should apply across all degree subjects [8]. However, other studies show that not all assessment methods suit different programmes or even courses [9]. Thus, student assessment methods in HE can be generally divided into two main categories: Exam-based assessments, which include different forms such as closed and open book examinations, essay-type exams, multiple choice exams, etc. and coursework-based assessments, which include research projects, assignments, etc. Different studies proved that students tend to gain higher marks from coursework-based assignments than they do from examinations [4]. Authors in [10] found that combining exam-based and coursework-based assessment methods produced up to 12% higher average marks than did examinations alone. As a result, this paper tries to take into account the different assessment methods of universities' modules and their effect on the students' academic performance, and also formulates the differences in assessment methods into a new attribute that can be part of further data mining processes.

## III. STUDENT TRANSCRIPT DATA

In this study, 4662 modules with different assessment methods that were collected from a UK University, are investigated. Basic average calculations of module marks in various departments (Business, Computer Science, Math, Electrical and Computer Systems Engineering, Civil Engineering, and Mechanical Engineering) are calculated for students taking modules with exam-based, coursework-based or a mixture of both assessment methods. A simple t-test was applied on the data (Table I) in order to measure the difference between means of each pair of variables. Results show that there is a statistically significant difference between the exam-based and coursework-based assessments (with 95% confidence level (which equates to declaring statistical significance at the $p<0.05$ level, a t-value of -5.06 and a P-value of 0.001).

TABLE I.    AVERAGE MARKS BASED ON ASSESSMENT METHOD

| Department | Student number | Average module mark of students | | |
|---|---|---|---|---|
| | | Exam-based assessment | Coursework-based assessment | Mixed assessment |
| Business | 54960 | 59.77 | 60.83 | 60.01 |
| Civil Engineering | 34892 | 58.78 | 63.74 | 60.70 |
| Computer Science | 19800 | 58.18 | 64.40 | 58.87 |
| Electronic and Computer Systems Engineering | 13740 | 59.55 | 63.26 | 57.00 |
| Math | 24152 | 61.59 | 66.00 | 61.17 |
| Mechanical Engineering | 31385 | 58.80 | 64.26 | 60.24 |

Applying the t-test to measure the significance of difference between each pair of variables Ex-CW, Ex-Mix, and CW-Mix assessment methods results in Table II. As shown in Table II, the p-value of the t-test between the fields: exam-based and coursework-based assessments (0.002) and both exam and coursework-based assessments, and coursework-based assessment (0.004) is less than 0.05, which indicates the statistical significant difference between these assessment methods. On the other hand, there seems to be no statistical significant difference between the exam-based assessment and the mixture of both exam and coursework assessment methods, since the P-Value was 0.749 which is greater than 0.05.

TABLE II.    P-VALUES OF T-TEST

| Assessment method | p-value | t-value |
|---|---|---|
| Exam and coursework | 0.002 | -4.5 |
| Mixed and exam | 0.749 | 0.39 |
| Mixed and coursework | 0.004 | -3.99 |

Throughout the literature, module assessment methods in HE have been investigated. Various studies such as [4] show that students tend to get higher marks when assessed using coursework than when assessed using exam-based assessment. Table I shows that this study initially does not contradict with previous studies, by confirming that students who are assessed using coursework tend to get higher marks than those who are assessed using exams or a mixture of both coursework and exams. The next section explains educational data used and their attributes before processing the data and applying data mining techniques.

### A. Understanding Student Transcript Data

Student transcript data were collected, and they consisted of files with hundreds of thousands of records. The standard DM practice suggests that the data in these files had to be first understood, then cleaned, and finally the most significant factors had to be highlighted in order to further process these data using EDM techniques. Normally, educational data are discrete, i.e. either numeric or categorical data, and noise-free. The lack of noise in educational data is due to the fact that there are not measured. They are either collected automatically or checked carefully [8]. On the other hand, missing data values exist usually in the cases where students – for example - skip answering a given questionnaire or in other cases where teachers skip checking attendance. Humans normally do this type of errors, referred as data entry errors [11]. The investigated data represent around 230,823 student records representing 19,886 students in a total of six departments of a UK University. Each one of these department data sheets contains a number of student records. For each record, a number of attributes that represent a student's academic accomplishments during three levels (preparatory, 1st, and 2nd levels) are divided as:

- Student-Related Attributes: These attributes highlight the status of the students, including: (i) Module Mark: Student's mark in a certain module, (ii) Exam Mark: The mark achieved by a student on the exam-based assessment part, and (iii) Cswk Mark: The mark achieved by a student on the coursework-based assessment part.

- Module-Related Attributes: A group of attributes that describe a certain module and its characteristics including: Coursework Weighting (CWW): Alternatively, this attribute indicates the ratio of coursework-based assessment to the total mark of a module.

For each module, the values of these two attributes complement each other to reach maximum total mark (i.e., 100). For example, if the Exam Weighting for a certain module is 75, the Coursework Weighting must be 25 in order to reach 100, and so on. For each department, different ratios between exam weighting and coursework weighting will be described in detail.

### B. Causes of Errors in Educational Data

Prior to storing educational data, they are normally processed by the involvement of human interaction, or computation, or both. Sources of errors in databases are categorized into four main types: data entry errors, measurement errors, distillation errors, and data integration errors [11]. As mentioned before, since educational data are not measured using instruments, the errors are caused by humans (i.e. through data entry errors) so these data have minimum, or zero noise compared to other non-educational data types. The following section will present the pre-processed students' transcript data in a different way in order to increase the data mining accuracy by extracting a new factor from the assessment method and use it to refine student marks to ensure more accuracy prior to processing them using EDM.

### IV. COURSEWORK ASSESSMENT RATIO (CAR)

In order to refine student's marks based on modules' assessment methods, the data have been processed through various stages. Each of these stages enhances the data in terms of readability by statistical software and ability to extract knowledge easily. The first step is to categorize the CW to EX ratios. The categorization algorithm relies on the number of classes that the ratio between CW to EX weighting can have based on ratios of CW to EX the original data have. Namely, each department has its own classification of CW to EX weighting ratios. CAR represents the ratio of CW weighting for each module, which by default will reflect the ratio of EX weighting. Table III shows the different classes:

TABLE III.     CLASSES OF CW TO EX WEIGHTING RATIOS

| CW | Model assessment method | | | | | | | | | | | | | | | | |
|---|---|---|---|---|---|---|---|---|---|---|---|---|---|---|---|---|---|
| | 0 | 10 | 15 | 20 | 25 | 30 | 35 | 40 | 45 | 50 | 55 | 60 | 65 | 66 | 70 | 75 | 100 |
| Business | ✓ | ✓ | | ✓ | ✓ | ✓ | | ✓ | | ✓ | | ✓ | ✓ | | | | ✓ |
| CEng | ✓ | ✓ | ✓ | ✓ | ✓ | ✓ | | ✓ | ✓ | ✓ | | | | | | ✓ | ✓ |
| CS | ✓ | ✓ | | ✓ | ✓ | ✓ | | ✓ | | ✓ | ✓ | ✓ | ✓ | | ✓ | | ✓ |
| ECSEng | ✓ | ✓ | ✓ | ✓ | ✓ | ✓ | ✓ | ✓ | | ✓ | | ✓ | | ✓ | | | ✓ |
| Math | ✓ | ✓ | | ✓ | ✓ | ✓ | | ✓ | | ✓ | | | | | ✓ | | ✓ |
| MEng | ✓ | ✓ | | ✓ | ✓ | ✓ | | ✓ | | ✓ | | | | | ✓ | ✓ | ✓ |

Table III shows that there is no department that shares the same ratio classes with the other departments, i.e. each department has its own unique ratios between CW to EX. Thus, filling the missing values in the Table is not a solution since doing so yields incorrect data. This research considers the department with the greatest number of classes to start with, and then generalizes the findings on the other departments while bearing in mind the change of ratios. Two departments have 12 complete ratios (the CS and ECSEng departments). In this paper, the CS department was chosen for further processing.

### V. REFINING STUDENTS MARKS BASED ON CAR

In order to obtain the relation between the student's module mark and CAR which represents the CWW to EXW ratios, simple quadratic regression was used. Regression analysis is being used to infer relationships between the independent (CAR) and dependent (refined module mark) variables. The variable of CW ratio was used, since the simple quadratic regression is more suitable for one variable relation. The choice of choosing quadratic over linear is based on the R-squared (coefficient of determination): when the R-squared is higher, then the model fits data more. For the case of quadratic regression, the coefficient is 2.90%, which is higher than in the linear model (2.77%). By applying simple quadratic regression on the data with CAR as a variable for module mark as a response, we achieved the following fitted regression line:

$$RMM = MM - 12.77(CAR) + 5.873(CAR)^2 \qquad (1)$$

where $RMM$ is the refined module mark after fitting and $MM$ is the current module mark.

By applying (1) on student transcript data of the CS Department, an additional field will be added which contains the $RMM$ for each student at the department. $RMM$s are self-explained when referring to Table I that compares the average module marks for student attending modules according to different assessment methods. The more the percentage of $CWW$, the more the added marks to $MM$, and vice versa. Any EDM that considers student marks with different assessment methods should consider adding a sub-process within the data pre-processing phase that takes into account the difference between those assessment methods. Figure 1 shows the addition of sub-processes 2-5 on raw students' module marks. The task of these sub-processes is to take the differences in assessment methods into account. These sub-processes are utilized because applying conventional DM processes on educational data may not produce accurate results.

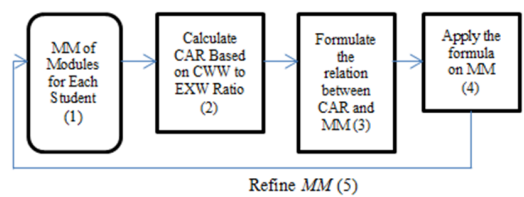

Fig. 1.     Refining students' module marks sub-process

### VI. THE EFFECT OF CAR ON DM'S PREDICTION ACCURACY

In order to verify the effect of the newly constructed variable, prediction of a student's third year average mark using his first and second year's average was compared to predicted third year's average using first and second year's average and the newly constructed variable, CAR. This comparison may highlight how accurate the prediction process can be by adding new attributes that reflect the nature of the module. Orange Canvas Tool is a simple data analysis tool with clever data visualization and interactive data exploration for rapid qualitative analysis with clean visualizations. Moreover, there are many prediction techniques which can be investigated using the Orange Data Mining Tool, but one of the most popular technique that could be used is Random Forest [12].

The Random Forest was used in this study to evaluate the role of CAR in increasing the prediction accuracy. The data of 406 undergraduate students of the Computer Science Department included their first, second and third years' average marks, and their average CAR for all years. The comparison was done including and excluding CAR as an input attribute. Figures 3 and 4 show the confusion matrices of the comparisons and how CAR affected the accuracy of predicting students' third year average marks.

TABLE IV.     PREDICTING THIRD YEAR'S AVERAGE FROM FIRST AND SECOND YEAR'S EXCLUDING CAR

|  |  | Prediction | | | | | | |
|---|---|---|---|---|---|---|---|---|
|  |  | Fail | First | Lower second | Pass | Third | Upper second |  |
| Correct Class | Fail | 0 | 1 | 3 | 0 | 5 | 3 | 12 |
|  | First | 0 | 36 | 0 | 0 | 0 | 37 | 73 |
|  | Lower Second | 0 | 0 | 0 | 0 | 6 | 31 | 47 |
|  | Pass | 0 | 0 | 0 | 0 | 0 | 3 | 3 |
|  | Third | 0 | 0 | 8 | 0 | 9 | 13 | 30 |
|  | Upper Second | 0 | 5 | 4 | 0 | 0 | 110 | 119 |
|  |  | 0 | 42 | 25 | 0 | 20 | 197 | 284 |

TABLE V.     PREDICTING THIRD YEAR'S AVERAGE FROM FIRST AND SECOND YEAR'S INCLUDING CAR AS AN ATTRIBUTE

|  |  | Prediction | | | | | | |
|---|---|---|---|---|---|---|---|---|
|  |  | Fail | First | Lower second | Pass | Third | Upper second |  |
| Correct Class | Fail | 3 | 2 | 6 | 0 | 0 | 2 | 13 |
|  | First | 0 | 59 | 0 | 0 | 0 | 23 | 82 |
|  | Lower Second | 0 | 4 | 16 | 0 | 0 | 26 | 46 |
|  | Pass | 0 | 0 | 1 | 0 | 0 | 2 | 3 |
|  | Third | 1 | 1 | 14 | 0 | 0 | 9 | 25 |
|  | Upper Second | 0 | 12 | 4 | 0 | 0 | 99 | 115 |
|  |  | 4 | 78 | 41 | 0 | 0 | 161 | 284 |

The ROC (receiver operating characteristics) curve has been introduced to evaluate ranking performance of machine learning algorithms [13]. Author in [14] has compared popular machine learning algorithms using the area under the curve (AUC) that represents the proportion of false positive rate covered by the curve of true positive rate and found that AUC exhibits several desirable properties compared to classification accuracy (CA). Table V shows an enhancement on the prediction probabilities for each of the mark classes (Fail, Pass, Third, Lower second, Upper second, and First classes) compared to probabilities of prediction shown in Table IV. The Random Forest scored the highest AUC when CAR was considered (Table V) with AUC value of 0.9304 and 0.0696 error rate. The AUC represents the proportion of false positive rate covered by the curve of true positive rate. In other words, the bigger the area of the curve, the more items are classified successfully as presumed, and with the increase of AUC, the mean profit difference also increases. When CAR was excluded as an input (Table IV), the AUC was lower with value 0.9073 and the error rate was 0.0927.

## VII.  RESULTS AND DISCUSSION

This study shows that the pre-processing phase of educational data should include additional sub-phases that deal, not only with noise or missing data, but also with data refinement so as to cope with differences between various educational systems and their data. Therefore, the attribute CAR was constructed in order to take the different modules' assessment methods into account while preparing student transcript data. The effect of CAR on the prediction process using Random Forest classification was investigated and applied on the equation of RMM. It was shown that considering RMM increases the accuracy of predicting students' marks. By refining students' marks, they either increase or decrease depending on the ratio between CWW to EXW for each student during his study. For instance, by considering a student x as an example, the student had enrolled in 32 modules during his/her study at the Department of CS. Nineteen out of 32 modules are 100% exam-based assessed modules, 7 are assessed by a mixture of coursework and examination, while only 6 modules are 100% assessed by coursework only. Despite that the majority of the modules are assessed through examination only, which means that the student gets no extra marks compared with coursework-based modules, the rest of the modules give the student extra marks and hence add to his overall average. In numbers, 19 modules have 0 CAR value, which means that *RMM=MM*. On the other hand, the rest of the modules have values of CAR ranging from 0.1 to 1, which means that the *RMM* is always less than *MM* for those courses. This decrement in the marks is due to the fact that students get higher marks in modules that are assessed by coursework or mixture of coursework and examinations. Therefore, to balance the module marks and the overall average, the formula decrements the module marks by varying percentage depending on the CWW to EXW to ratios. Table VI shows the differences in module marks and overall average for the example student.

TABLE VI.    REFINING STUDENT X MARKS BASED ON ENROLLED MODULES' ASSESSMENT METHODS

|  | Average MM | Average RMM |
|---|---|---|
| **19 exam-based modules** | 48.6 | 48.6 |
| **6 coursework-based modules** | 60.3 | 52.7 |
| **7 mixed EX and CW modules** | 60.4 | 58.3 |
| **Total 32 modules** | 56.4 | 53.2 |

By following the procedure in Figure 1 on a real student's marks, Table VI shows that *RMM* remains unchanged for the student when the assessment method of the enrolled modules is purely exam-based. Alternatively, when the assessment method of the enrolled modules is purely coursework-based, the *RMM* is refined down on average 7.6 compared to *MM*. Finally, a mixture of EX and CW based modules yields less refinement of the *RMM* (2.1 marks) compared to *MM* for the same student. This means that student in the example who is taking 32 modules of different types may have his average marks refined down by 3.2 when applying the proposed procedure. Upon deriving CAR and examining its effect on students' marks, Random Forest prediction technique was used to validate the enhancement on AUC. It was shown that by including MAR as an input for predicting students' third year from their first and second years, AUC was enhanced from 0.9073 to 0.9304. The derived CAR enhances the predictability of students' third year marks from their first two years, which means that additional preparation steps, such as this paper's module mark refinement based on modules' assessment methods, are required on the student transcript data prior to applying DM techniques for improved predictability. Additionally, the findings of this paper will be generalized in different departments and Universities that may have various assessment methods, in other words, the refinement procedure should be considered as a sub-process within the EDM data pre-processing phase when dealing with different modules and their marks.

## VIII.   CONCLUSIONS AND FUTURE WORK

During the last few decades, there has been an increased interest in coursework-based assessment in the UK and other countries due to various educational and personal arguments such as its learning effectiveness, the lack of time limits and stress, etc. This increased interest led to discovering that students who are assessed by coursework tend to achieve higher marks than those who are assessed by examinations in the same modules. However, it seems that no studies have considered this increase in marks in the data pre-processing phase. More generally, this led to a conclusion that applying conventional DM processes on educational data may not produce accurate results. In this paper, a model that refines students' marks based on enrolled modules' assessment methods has been proposed. The model represents a sub-process through which module assessment methods are considered for further processing using a new attribute that reflects the ratio of coursework weightings. The key to refine students' marks is to develop a ratio that represents the ratio of CW weighting for each module, which by default will reflect the ratio of EX weighting. This ratio, coursework assessment ratio (CAR), has been extracted using simple quadratic regression on the MM (module marks) variable. Based on CAR values, *RMM* (refined module marks) were calculated and compared to the original *MM*. Although this increase on marks was proved in literature and throughout this research, none of the previous studies considered this increase as a feedback to the data pre-processing phase of the EDM, which has mostly been following conventional DM pre-processing methods while neglecting the differences between the types of data being processed. Therefore, it is vital to consider this feedback generation for EDM pre-processing phase. This means that the formulation of the relation between CAR and the current *MM* should be part of the EDM pre-processing phase. This study investigated the effect of CAR on the prediction process using Random Forest classification technique. It was shown that considering CAR as an attribute increases the accuracy of predicting students' third year averages based on their first and second year's averages. For future work, the predictability of individual module results based on their assessment method and further investigation into the group course work can be applied to improve predictability.